\documentclass[12pt]{article}
\usepackage{amsmath,amssymb, amsfonts, amsthm,amscd}
\usepackage[T2A]{fontenc}
\usepackage[cp1251]{inputenc}
\usepackage[russian,ukrainian,english]{babel}
\usepackage{geometry}

\newcommand{\case}[1]{\textbf{Case $\mathbf{#1}$.}\ }

\newtheorem{Theorem}{Theorem}
\newtheorem{Corollary}{Corollary}

\newtheorem{Lemma}{Lemma}

\newenvironment{LemmaProof}{\textbf{Proof. }}{\par\noindent\textbf{The Lemma is proved.}}
\newenvironment{TheoremProof}{\textbf{Proof. }}{\par\noindent\textbf{The Theorem is proved.}}

\title{{\Large \textbf{Interval colorings of complete bipartite graphs and trees}}}

\author{\normalsize \textbf{R.R. Kamalian}}

\date{\small{\small{A translation from Russian of the work of R.R. Kamalian "Interval colorings of complete
bipartite graphs and trees", Preprint of the Computing Centre of the
Academy of Sciences of Armenia, Yerevan, 1989. (Was published by the
decision of the Academic Council of the Computing Centre of the
Academy of Sciences of Armenian SSR and Yerevan State University
from 7.09.1989)}}}

\begin{document}

\maketitle

\bigskip

In the work interval colorings \cite{Oranj} of complete bipartite
graphs and trees are investigated. The obtained results were
announced in \cite{Gorki}. Non defined concepts can be found in
\cite{Zikov, Harary}.

Let $G=(V(G),E(G))$ be an undirected graph without multiple edges
and loops. The degree of a vertex $x$ in $G$ is denoted by $d_G(x)$,
the greatest degree of vertices -- by $\Delta(G)$, the chromatic
index of $G$ -- by $\chi'(G)$.

Interval $t$-coloring of a graph $G$ is a proper coloring of edges
of $G$ by the colors $1,\ldots,t$, at which by each color $i$,
$1\leq i\leq t$, at least one edge $e_i\in E(G)$ is colored, and
edges incident with each vertex $x\in V(G)$ are colored by $d_G(x)$
consecutive colors.

A graph $G$ is called interval colorable if there is $t\geq1$ for
which $G$ has an interval $t$-coloring. For an interval colorable
graph $G$, we denote by $w(G)$ and $W(G)$, respectively, the least
and the greatest value of $t$, for which $G$ has an interval
$t$-coloring.

If $\alpha$ is a proper edge coloring of a graph $G$, then the color
of an edge $e\in E(G)$ at this coloring is denoted by $\alpha(e,G)$
or, if it is clear which graph is spoken about, by $\alpha(e)$.

Let $k$ and $l$ be positive integers. Let us denote by $\sigma(k,l)$
the greatest common divisor of $k$ and $l$. The algorithm of Euclid
for finding of $\sigma(k,l)$ consists of the construction of
sequences $(F_i(k,l))$, $(f_i(k,l))$, $i=1,2,\ldots$, defined as
follows: $F_1(k,l)=\max\{k,l\}$, $f_1(k,l)=\min\{k,l\}$; if
$F_1(k,l)=f_1(k,l)$ then the construction of the sequences is
finished, and if $F_1(k,l)>f_1(k,l)$ then
$F_{i+1}(k,l)=\max\{F_i(k,l)-f_i(k,l),f_i(k,l)\}$,
$f_{i+1}(k,l)=\min\{F_i(k,l)-f_i(k,l),f_i(k,l)\}$, $i=1,2,\ldots$.
The algorithm is completed at the finding of such $j$ (let us denote
it by $s(k,l)$) for which $F_j(k,l)=f_j(k,l)=\sigma(k,l)$.

Let $H(\mu,\nu)$ be a $(0,1)$-matrix with $\mu$ rows, $\nu$ columns,
and with elements $h_{ij}$, $1\leq i\leq\mu$, $1\leq j\leq \nu$. The
$i$-th row of the matrix $H(\mu,\nu)$, $1\leq i\leq\mu$, is called
collected, if $h_{ip}=h_{iq}=1$, $p\leq t\leq q$ imply $h_{it}=1$,
and the inequality $\sum_{j=1}^{\nu}h_{ij}\geq1$ holds. Similarly,
the $j$-th column of the matrix $H(\mu,\nu)$, $1\leq j\leq\nu$, is
called collected, if $h_{pj}=h_{qj}=1$, $p\leq t\leq q$ imply
$h_{tj}=1$, and the inequality $\sum_{i=1}^{\mu}h_{ij}\geq1$ holds.
For the $i$-th row of the matrix $H(\mu,\nu)$, all rows and columns
of which are collected, define a number
$\varepsilon(i,H(\mu,\nu))=\min_{h_{ij}=1}j$, $i=1,\ldots,\mu$. For
the $j$-th column of the matrix $H(\mu,\nu)$, all rows and columns
of which are collected, define a number
$\xi(j,H(\mu,\nu))=|\{i/\;\varepsilon(i,H(\mu,\nu))=j,1\leq i\leq
\mu\}|$, $j=1,\ldots,\nu$. $H(\mu,\nu)$ is called an $r$-regular
$(r\geq1)$ matrix, if $\sum_{j=1}^{\nu}h_{ij}=r$, $i=1,\ldots,\mu$.
$H(\mu,\nu)$ is called a collected matrix, if all its rows and
columns are collected, $h_{11}=h_{\mu\nu}=1$, and the inequality
$\varepsilon(1,H(\mu,\nu))\leq\ldots\leq\varepsilon(\mu,H(\mu,\nu))$
holds. $(0,1)$-matrices $A(\alpha,\gamma)$ and $B(\beta,\gamma)$
with elements $a_{ij}$, $1\leq i\leq\alpha$, $1\leq j\leq\gamma$ and
$b_{ij}$, $1\leq i\leq\beta$, $1\leq j\leq\gamma$, respectively, are
called equivalent, if
$\sum_{i=1}^{\alpha}a_{ij}=\sum_{i=1}^{\beta}b_{ij}$,
$j=1,\ldots,\gamma$. An $r'$-regular $(r'\geq1)$ matrix
$H'(\mu',\nu')$ and an $r''$-regular $(r''\geq1)$ matrix
$H''(\mu'',\nu'')$ are called mutually conformed, if $r'=\mu''$ and
$r''=\mu'$.

\begin{Lemma}\label{lem1}
If a collected $n$-regular $(n\geq1)$ matrix $P(m,w)$ with elements
$p_{ij}$, $1\leq i\leq m$, $1\leq j\leq w$ is equivalent to a
collected $m$-regular $(m\geq1)$ matrix $Q(n,w)$ with elements
$q_{ij}$, $1\leq i\leq n$, $1\leq j\leq w$, then $w\geq
m+n-\sigma(m,n)$.
\end{Lemma}

\textbf{Proof} by induction on $s(m,n)$. If $s(m,n)=1$, then
$m=n=\sigma(m,n)$, and, clearly, the lemma is true. Let
\begin{equation}
s(m,n)=z_0>1 \label{eq:1}
\end{equation}
and the lemma is supposed
to be true for mutually conformed equivalent an $n'$-regular
$(n'\geq1)$ matrix and an $m'$-regular $(m'\geq1)$ matrix with
$s(m',n')<z_0$. Assume, in opposite to the desired, that
\begin{equation}
w<m+n-\sigma(m,n) \label{eq:2}
\end{equation}
and, for definition,
\begin{equation}
m\geq n \label{eq:3}
\end{equation}

Let us note that $\varepsilon(n,Q(n,w))+m-1=w<m+n-\sigma(m,n)\leq
m+n-1$, which implies
\begin{equation}
\varepsilon(n,Q(n,w))<n \label{eq:4}
\end{equation}

From (\ref{eq:3}) we conclude:
\begin{equation}
\sum_{i=1}^m p_{ij}=\sum_{r=1}^j\xi(r,P(m,w)),\quad j=1,\ldots,n
\label{eq:5}
\end{equation}
\begin{equation}
\sum_{i=1}^n q_{ij}=\sum_{r=1}^j\xi(r,Q(n,w)),\quad j=1,\ldots,n
\label{eq:6}
\end{equation}

From the equivalence of the matrices $P(m,w)$ and $Q(n,w)$, and from
the relations (\ref{eq:3}) -- (\ref{eq:6}), it follows that:
\begin{equation}
\xi(j,P(m,w))=\xi(j,Q(n,w)),\quad j=1,\ldots,n \label{eq:7}
\end{equation}

\begin{equation}
\sum_{i=1}^m p_{in}=\sum_{i=1}^n q_{in}=n \label{eq:8}
\end{equation}

Let us form from matrices $P(m,w)$ and $Q(n,w)$, respectively,
matrices $P_1(m-n,w-n)$ and $Q_1(n,w-n)$ of smaller dimensions by
the following way: form $P_1(m-n,w-n)$ from $P(m,w)$ by removing
that and only that elements $p_{ij}$, for which at least one of the
inequalities $i\leq n$, $j\leq n$ holds; form $Q_1(n,w-n)$ from
$Q(n,w)$ by zeroing that and only that elements $q_{ij}$, for which
$j<\varepsilon(i,Q(n,w))+n$, and further removing of all elements of
first $n$ columns.

From (\ref{eq:1}) and (\ref{eq:3}) it follows that
\begin{equation}
m>n \label{eq:9}
\end{equation}

From the construction of matrices $P_1(m-n,w-n)$, $Q_1(n,w-n)$ and
from the relations (\ref{eq:1}), (\ref{eq:3}), (\ref{eq:7}),
(\ref{eq:8}) it follows that $P_1(m-n,w-n)$ is a collected
$n$-regular $(n\geq1)$ matrix, $Q_1(n,w-n)$ is an equivalent to it
collected $(m-n)$-regular $(m-n\geq1)$ matrix. Clearly,
$P_1(m-n,w-n)$ and $Q_1(n,w-n)$ are mutually conformed,
$s(m-n,n)<z_0$. From here, by the assumption of induction, we have
the inequality $w-n\geq(m-n)+n-\sigma(m-n,n)$, or
\begin{equation}
w\geq m+n-\sigma(m-n,n) \label{eq:10}
\end{equation}

From (\ref{eq:9}) we conclude $\sigma(m-n,n)=\sigma(m,n)$, and,
taking (\ref{eq:10}) into account, we obtain the inequality $w\geq
m+n-\sigma(m,n)$, which contradicts the assumption (\ref{eq:2}).

\textbf{The Lemma is proved.}

Let $K_{m,n}$ be a complete bipartite graph with the set
$V(K_{m,n})=\{x_1,\ldots,x_m,y_1,\ldots,y_n\}$ of vertices and the
set $E(K_{m,n})=\{(x_i,y_j)/\;1\leq i\leq m, 1\leq j\leq n\}$ of
edges.

\begin{Lemma}\label{lem2}
For arbitrary positive integers $m$ and $n$, $K_{m,n}$ has an
interval $(m+n-1)$-coloring.
\end{Lemma}

\begin{LemmaProof}
For obtaining of an interval $(m+n-1)$-coloring of the graph
$K_{m,n}$, color the edge $(x_i,y_j)$, $1\leq i\leq m$, $1\leq j\leq
n$, by the color $i+j-1$.
\end{LemmaProof}

\begin{Theorem}\label{thm1}
For arbitrary positive integers $m$ and $n$,
\begin{enumerate}
  \item $K_{m,n}$ is interval colorable,
  \item $w(K_{m,n})=m+n-\sigma(m,n)$,
  \item $W(K_{m,n})=m+n-1$,
  \item if $w(K_{m,n})\leq t\leq W(K_{m,n})$, then $K_{m,n}$ has an
  interval $t$-coloring.
\end{enumerate}
\end{Theorem}

\begin{TheoremProof}
The proposition 1) of the theorem immediately follows from the lemma
\ref{lem2}. From the already proved proposition 1) and from the
corollary of the theorem 1 of the work \cite{Oranj} we have
$W(K_{m,n})\leq|V(K_{m,n})|-1=m+n-1$. From here and from the lemma
\ref{lem2} the proposition 3) of the theorem follows.

Now let us be convinced of $w(K_{m,n})\geq m+n-\sigma(m,n)$.
Consider an interval $w(K_{m,n})$-coloring of the graph $K_{m,n}$.
For $v\in V(K_{m,n})$, let us denote by $\lambda(v)$ the least among
colors of edges incident with $v$. Clearly, without loss of
generality, we can assume that
\begin{equation}
\lambda(x_1)\leq\dots\leq\lambda(x_m);\quad
\lambda(y_1)\leq\dots\leq\lambda(y_n) \label{eq:11}
\end{equation}

Define a matrix $X=(x_{ij})$ with $m$ rows and $w(K_{m,n})$ columns:
$$
x_{ij}=\left\{
\begin{array}{ll}
1, & \textrm{if there is an edge colored by $\;j$ incident with the vertex $\;x_i$}\\
0 & \textrm{-- otherwise},\\
\end{array}
\right.
$$
$1\leq i\leq m$, $1\leq j\leq w(K_{m,n})$.

Define a matrix $Y=(y_{ij})$ with $n$ rows and $w(K_{m,n})$ columns:
$$
y_{ij}=\left\{
\begin{array}{ll}
1, & \textrm{if there is an edge colored by $\;j$ incident with the vertex $\;y_i$}\\
0 & \textrm{-- otherwise},\\
\end{array}
\right.
$$
$1\leq i\leq n$, $1\leq j\leq w(K_{m,n})$.

From properties of the considered coloring and inequalities
(\ref{eq:11}) it follows that $X$ is a $n$-regular $(n\geq1)$
collected matrix, and $Y$ is an equivalent to it $m$-regular
$(m\geq1)$ collected matrix. It is also clear that $X$ and $Y$ are
mutually conformed. It follows from the lemma \ref{lem1} that
$w(K_{m,n})\geq m+n-\sigma(m,n)$.

Evidently, for the completion of the proof of the theorem it is
suffice to show, that, if $m+n-\sigma(m,n)\leq t\leq m+n-1$, then
$K_{m,n}$ has an interval $t$-coloring.

Let $t=m+n-\sigma(m,n)+\mu$, where
\begin{equation}
0\leq\mu\leq\sigma(m,n)-1 \label{eq:12}
\end{equation}

Let us denote by $G_1$ the subgraph of the graph $K_{m,n}$ induced
by the vertices
$x_1,\dots,x_{\sigma(m,n)},y_1,\dots,y_{\sigma(m,n)}$.

Let $p=\frac{m}{\sigma(m,n)}$, $q=\frac{n}{\sigma(m,n)}$.

$G_1$ is a regular complete bipartite graph. From the proposition 2
of the work \cite{Oranj} it follows that
\begin{equation}
\chi'(G_1)=\Delta(G_1)=w(G_1)=\sigma(m,n) \label{eq:13}
\end{equation}

From the already proved proposition 3) of the theorem we have
\begin{equation}
W(G_1)=2\sigma(m,n)-1 \label{eq:14}
\end{equation}

From the relations (\ref{eq:12}) -- (\ref{eq:14}) we obtain
\begin{equation}
\Delta(G_1)=w(G_1)\leq\sigma(m,n)+\mu\leq W(G_1) \label{eq:15}
\end{equation}
Since $G_1$ is a regular graph then from (\ref{eq:13}),
(\ref{eq:15}) and the proposition 2 of the work \cite{Oranj} it
follows that there exists an interval $(\sigma(m,n)+\mu)$-coloring
$\alpha$ of the graph $G_1$. Now, in order to receive an interval
$t$-coloring of the graph $K_{m,n}$, it is suffice for
$\tau=1,\ldots, p-1$ and $\varepsilon=1,\ldots, q-1$ to color the
edge $(x_{i+\tau\sigma(m,n)},y_{j+\varepsilon\sigma(m,n)})$ of the
graph $K_{m,n}$ by the color
$(\tau+\varepsilon)\cdot\sigma(m,n)+\alpha((x_i,y_j),G_1)$, $1\leq
i\leq\sigma(m,n)$, $1\leq j\leq\sigma(m,n)$.
\end{TheoremProof}

\begin{Corollary}\label{cor1}
If $\sigma(m,n)=1$, then $K_{m,n}$ has an interval $t$-coloring iff
$t=m+n-1$.

Let $D$ be a tree, $V(D)=\{b_1,\ldots,b_{\beta}\}$, $\beta\geq1$.
Let us denote by $L(b_i,b_j)$ the path connecting the vertices $b_i$
and $b_j$, by $VL(b_i,b_j)$ and $EL(b_i,b_j)$ -- the sets of
vertices and edges of this path, respectively, $1\leq i\leq\beta$,
$1\leq j\leq\beta$. For the path $L(b_i,b_j)$, $1\leq i\leq\beta$,
$1\leq j\leq\beta$, let us introduce a notation:
$$
ML(b_i,b_j)=|EL(b_i,b_j)|+|\{(x,y)/\;(x,y)\in E(D),x\in
VL(b_i,b_j),y\not\in VL(b_i,b_j)\}|.
$$

Let
$$
M(D)=\max_{1\leq i\leq\beta,1\leq j\leq\beta}ML(b_i,b_j).
$$
\end{Corollary}

\begin{Lemma}\label{lem3}
If a tree $D$ is interval colorable, then $W(D)\leq M(D)$.
\end{Lemma}

\begin{LemmaProof}
Without loss of generality, we can assume that $|E(D)|>1$ (otherwise
the lemma is evident). Consider an interval $W(D)$-coloring $\alpha$
of the tree $D$. Let $\alpha(e_1)=1$, $\alpha(e_2)=W(D)$,
$e_1=(x',y')$, $e_2=(x'',y'')$. Without loss of generality we can
assume that $|EL(x',x'')|>|EL(y',y'')|$. Let us number the vertices
of the set $VL(x',x'')$ in the direction from $x'$ to $x''$:
$x'=z_0,z_1,\ldots,z_s,z_{s+1}=x''$, where $s\geq1$.

Let us note that
$\alpha((z_i,z_{i+1}))\leq1+\sum_{j=1}^i(d_D(z_j)-1)$,
$i=1,\ldots,s$. Consequently,
$W(D)=\alpha(e_2)=\alpha((z_s,z_{s+1}))\leq1+\sum_{j=1}^s(d_D(z_j)-1)=ML(x',x'')\leq
M(D)$.
\end{LemmaProof}

\begin{Lemma}\label{lem4}
If $D$ is a tree, and $\Delta(D)\leq t\leq M(D)$, then $D$ has an
interval $t$-coloring.
\end{Lemma}

\textbf{Proof} by induction on $|E(D)|$. If $|E(D)|=1$, then,
clearly, the lemma is true. Let $|E(D)|=k>1$, and assume that the
lemma is true for all trees $D'$ with $|E(D')|<k$.

\case{1} $M(D)<|E(D)|$.

In this case there is a pendent edge $e=(x,y)\in E(D)$, $d_D(x)=1$,
such, that its removing from $D$ gives a tree $D'$ with
$M(D')=M(D)$. Since $|E(D)|>1$, then $d_D(y)\neq1$. Clearly,
$d_{D'}(y)=d_D(y)-1$, $\Delta(D')\leq\Delta(D)$,
$|E(D')|=|E(D)|-1<k$, $\Delta(D')\leq t\leq M(D')$. By the
assumption of induction, there exists an interval $t$-coloring of
the tree $D'$. Suppose that the edges of $E(D')$ incident with the
vertex $y$ are colored in this coloring by the colors $\lambda_1(1),
\lambda_1(2),\ldots,\lambda_1(d_{D'}(y))$, where $1\leq\lambda_1(1)<
\ldots<\lambda_1(d_{D'}(y))\leq t$. If $\lambda_1(1)>1$, we shall
color the edge $e$ by the color $\lambda_1(1)-1$ and obtain an
interval $t$-coloring of the tree $D$. If $\lambda_1(1)=1$, then
$\lambda_1(d_{D'}(y))=d_{D'}(y)=d_{D}(y)-1$. We shall color the edge
$e$ by the color $d_D(y)$ and obtain an interval $t$-coloring of the
tree $D$.

\case{2} $M(D)=|E(D)|$.

\case{2a)} $t\leq M(D)-1$.

Let $e=(x,y)$ be a pendent edge in $D$, and $d_D(x)=1$. Since
$|E(D)|>1$, then $d_D(y)\neq1$. Let us denote by $D'$ the tree which
is obtained from the tree $D$ by removing of the edge $e$. Clearly,
$d_{D'}(y)=d_D(y)-1$, $\Delta(D')\leq\Delta(D)$, $M(D)-1\leq
M(D')\leq M(D)$, hence, $\Delta(D')\leq\Delta(D)\leq t\leq
M(D)-1\leq M(D')$. Since $|E(D')|=|E(D)|-1<k$, then, by the
assumption of induction, there exists an interval $t$-coloring of
the tree $D'$. Suppose that the edges of $E(D')$ incident with the
vertex $y$, are colored in this coloring by the colors
$\lambda_2(1), \lambda_2(2),\ldots,\lambda_2(d_{D'}(y))$, where
$1\leq\lambda_2(1)<\lambda_2(2)<\ldots<\lambda_2(d_{D'}(y))\leq t$.
If $\lambda_2(1)>1$, we shall color the edge $e$ by the color
$\lambda_2(1)-1$ and obtain an interval $t$-coloring of the tree
$D$. If $\lambda_2(1)=1$, then
$\lambda_2(d_{D'}(y))=d_{D'}(y)=d_{D}(y)-1$. We shall color the edge
$e$ by the color $d_D(y)$ and obtain an interval $t$-coloring of the
tree $D$.

\case{2b)} $t=M(D)$.

Clearly, without loss of generality, we can assume that
$ML(b_1,b_2)=M(D)$. Clearly, $d_D(b_1)=d_D(b_2)=1$. Let us number
the vertices of the path $L(b_1,b_2)$ in the direction from $b_1$ to
$b_2$: $b_1=z_0,z_1,\ldots,z_s,z_{s+1}=b_2$, where $s\geq1$. Let us
construct an interval $t$-coloring of the tree $D$. We shall color
the edge $(z_0,z_1)$ by the color $1$, the edge $(z_i,z_{i+1})$,
$i=1,\ldots,s$ -- by the color $1+\sum_{j=1}^i(d_D(z_j)-1)$.
$d_D(z_1)-2$ edges without a color incident with the vertex $z_1$,
will be colored by the colors $2,\ldots,d_D(z_1)-1$. $d_D(z_i)-2$
edges without a color incident with the vertex $z_i$,
$i=2,\ldots,s$, will be colored by the colors
$(1+\sum_{j=1}^{i-1}(d_D(z_j)-1))+1,\ldots,(1+\sum_{j=1}^i(d_D(z_j)-1))-1$.

\textbf{The Lemma is proved.}

From lemmas \ref{lem3} and \ref{lem4} we obtain

\begin{Theorem}
Let $D$ be a tree. Then
\begin{enumerate}
  \item $D$ is interval colorable,
  \item $w(D)=\Delta(D)$,
  \item $W(D)=M(D)$,
  \item if $w(D)\leq t\leq W(D)$, then $D$ has an
  interval $t$-coloring.
\end{enumerate}
\end{Theorem}

\bigskip

I thank A.S. Asratian for advices and attention to the work.

\end{document}